\documentclass
[aps,showpacs,showkeys,superscriptaddress]{revtex4}
\usepackage[cp1250]{inputenc}
\usepackage[T1]{fontenc}

\usepackage{graphicx}

\begin{document}

\newcommand{\rmi}[1]{_{\mathrm{#1}}}


\raggedbottom

\title{Optimal geometry for efficient loading of an optical dipole trap}

\author{Andrzej Szczepkowicz}

\affiliation{Institute of Experimental Physics, University of
Wrocław, Plac Maksa Borna 9, 50-204 Wrocław, Poland}

\author{Leszek Krzemień}

\author{Adam Wojciechowski}

\author{Krzysztof Brzozowski}

\affiliation{Institute of Physics, Jagiellonian University, Reymonta
4, 30-057 Kraków, Poland}

\author{Michael Kr{\"u}ger}

\affiliation{Institute of Physics, Jagiellonian University, Reymonta
4, 30-057 Kraków, Poland}

\affiliation{Max-Planck-Institute of Quantum
Optics, Hans-Kopfermann-Str.~1, 85748 Garching, Germany}

\author{Michał Zawada}

\affiliation{Institute of Physics, Nicolaus
Copernicus University, Grudziądzka 5, 87-100 Toruń, Poland}

\author{Marcin Witkowski}

\affiliation{Institute of Physics, University of Opole, Oleska 48,
45-052 Opole, Poland}

\author{Jerzy Zachorowski}

\author{Wojciech Gawlik}

\affiliation{Institute of Physics, Jagiellonian University, Reymonta
4, 30-057 Kraków, Poland}

\date{\today}

\begin{abstract}
One important factor which determines efficiency of loading cold
atoms into an optical dipole trap from a magneto-optical trap is
the distance between the trap centers. By studying this
efficiency for various optical trap depths (2--110~mK)
we find that for optimum dipole trap loading,
longitudinal displacements up to 15~mm are necessary. An
explanation for this observation is presented and compared with
other work and a simple analytical formula is derived for the
optimum distance between the trap centers.
\end{abstract}

\pacs%
{%
37.10.Gh,
37.10.De,
37.10.Vz
}


\maketitle

\section{Introduction\label{introduction}}

Optical dipole trapping is an established tool for storing,
manipulating and studying cold atomic gases
\cite{Chu1986,MetcalfStraaten1999,Grimm2000}. Optical dipole traps
(ODTs) helped to achieve Bose-Einstein condensate of several
elements \cite{optical-BEC-Rb, optical-BEC-Na, optical-BEC-Yb,
optical-BEC-Cr, optical-BEC-Cs} as well as to make ultra-stable
frequency standards \cite{clock}. ODTs are also widely
used for studies of various phenomena at quantum degeneracy, such
as the Mott insulator -- superfluid liquid transition
\cite{Mott-superfluid}.

The simplest realization of an optical dipole trap (ODT) is a
single, tightly focused Gaussian laser beam, tuned far below the
atomic resonance frequency. The trapping potential is proportional
to the light intensity which, for a single beam, results
in a highly elongated trap shape with optical potential minimum at
the trapping beam focus.

The ODT needs to be loaded by a precooled gas, and a standard way
of loading atoms into the ODT is to transfer them from a
magneto-optical trap (MOT). Kuppens \emph{et al.}
\cite{Kuppens2000} discuss ways of maximizing this transfer by
optimizing various parameters: MOT light intensities and
detunings, magnetic field gradient, the ODT depth, and alignment of
the traps. In particular, they found that controlling the geometry of the overlap of the MOT and ODT beams allows for substantial improvement of the transfer efficiency. They found that the loading rate is optimum with a longitudinal
displacement between the centers of the ODT and the MOT, and that
this optimum displacement increases with the ODT depth. Optimal
displacements between the ODT and MOT centers reported in Ref.
\cite{Kuppens2000} were about one-half of a MOT diameter (the
depths of the studied ODTs were 1--6 mK). These
observations were rather qualitative and not accompanied by
systematic quantitative analysis.

Since the loading of the trap is a crucial step in any
experimental work involving ODT, the purpose of our work was to
determine quantitatively the optimum displacement between the ODT
and MOT centers for maximum transfer of atoms in a wide range of
the ODT depths. The trap depths were changed by varying
the ODT laser power from 50 to 660 mW and detuning from 0.43 to
1.72 nm. The experimental setup and procedure are
described in the next Section, while the obtained results are
presented in Section \ref{results}, and in Section
\ref{discussion} we compare quantitatively our experimental
results with the models described by Kuppens \emph{et al.}
\cite{Kuppens2000} and O'Hara \emph{et al.} \cite{OHara2001}.

\section{Experiment}

\begin{figure}
\includegraphics{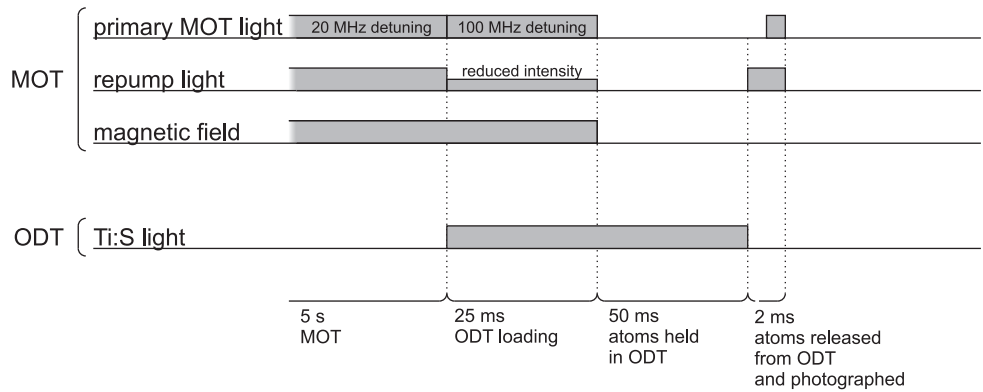}%
\caption%
{%
 \label{fig-timing}%
 The time sequence of loading the ODT from the MOT.
}
\end{figure}

We use a standard MOT setup with three pairs of orthogonal,
retroreflected beams for trapping $^{87}\mathrm{Rb}$ atoms. The
trapping light, red detuned from the $5^2{\mathrm S}_{1/2}\,F=2
\rightarrow 5^2{\mathrm P}_{3/2}\,F'=3$ transition, has a
total six-beam power of 15~mW. The repumping light (0.5 mW, tuned
to the $5^2{\mathrm S}_{1/2}\,F=1 \rightarrow 5^2{\mathrm
P}_{3/2}\,F'=2$ transition) is added to one pair of the trapping
beams. The radius of the beams is 8~mm ($1/e^2$ intensity). The
resulting MOT radius is 0.66~mm ($1/e^2$ density). After
collection, the atoms are further cooled by increasing the
detuning of the MOT beams to 100 MHz and decreasing the repumper
intensity (see Fig.~\ref{fig-timing}). At the end of the MOT
cooling stage the MOT temperature reaches $T\rmi{MOT}=17~\mu$K, as
determined from free expansion of the released atom cloud.

The dipole trap is formed by a focused beam of a
Ti:Sapphire laser. The ODT beam is first expanded, then focused
by a 25-cm focal length achromatic lens to a waist of
12~$\mu$m. The corresponding Rayleigh length is
$z\rmi{R}=0.54$~mm. We apply the ODT beam with powers,
$P\rmi{ODT}$, of 50, 100, 200, 400, and 660 mW, and red
detunings from the D$_2$ line, $\Delta\lambda\rmi{ODT}$, of 0.43,
0.86, and 1.72~nm. The resulting trap depths determined by the
ratio $P\rmi{ODT} / \Delta\lambda\rmi{ODT}$ were $U_0/k\rmi{B}$ =
2.1, 4.2, 8.4, 17, 33, 67, and 111 mK. For the trap-depth
estimation, a possible effect of the Rb D1 line has been
neglected as it does not exceed a few percent.

The transfer of the Rb atoms from the MOT to the ODT proceeded
with the timing shown in Fig.~\ref{fig-timing}; it is similar to
the sequence used by Kuppens \emph{et al.}
\cite{Kuppens2000}.

For each ODT laser power and detuning, the ODT-focus
position was gradually varied relative to the MOT center.
The described loading sequence (Fig.~\ref{fig-timing}) was
repeated for each position of the traps and the number of atoms
loaded into the ODT was monitored by integrating the fluorescence
image captured by the CCD camera.

\section{Results\label{results}}

\begin{figure}
\includegraphics{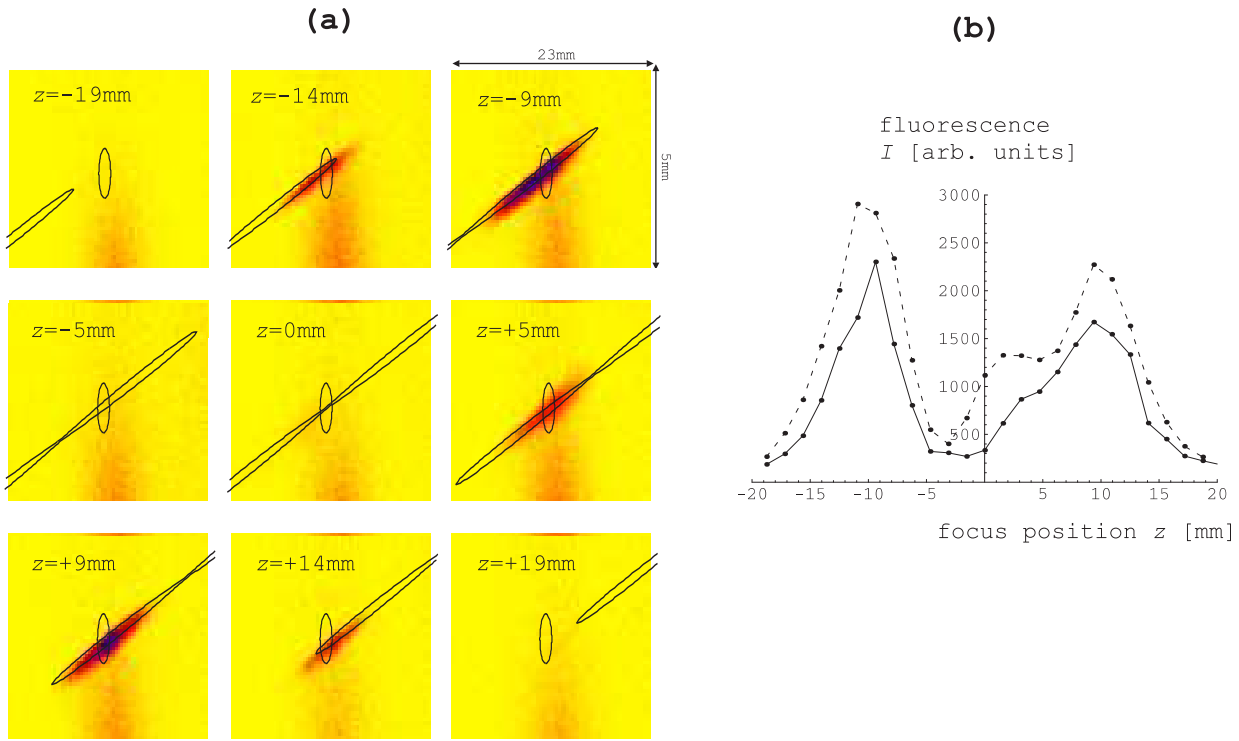}%
\caption%
{%
 \label{fig-33mK}%
 Loading of the 33-mK-deep dipole trap
 for different positions of the ODP laser focus
 with respect to the MOT center.
 (a)~Fluorescence images of atoms in the ODT,  vertically
 stretched for better visualization of the ODT potential.
 The elliptic contours mark the MOT positions
 (0.66 mm radius) before switching on the ODT.
 The elongated contours represent the ODT equipotential
 surfaces, $U=-2.5k\rmi{B}T\rmi{MOT}$.
 (b) Total fluorescence from the ODT, proportional to the number
 of atoms loaded into the ODT, versus the focus position.
 Solid line:  $P\rmi{ODT}=200$~mW, $\Delta\lambda$ =
 0.43~nm; dashed line: $P\rmi{ODT}=400$~mW, $\Delta\lambda$ =
 0.86~nm.
}
\end{figure}

Figure~\ref{fig-33mK} shows the results of the loading
measurements performed for the trap depth of $U_0/k\rmi{B}$ = 33
mK. The fluorescence images of the dipole trap are presented in
Fig.~\ref{fig-33mK}~(a) for several different positions $z$ of the
ODT relative to the MOT center. Superimposed are the
contours of the original MOT ($1/e^{2}$ density contour)
and the dipole trap (the equipotential line at
$U=-2.5k\rmi{B}T\rmi{MOT}$). The diffused clouds seen below the
trap contours are the atoms which were not captured into the ODT,
falling in the direction of gravity after MOT has been switched
off. Figure~\ref{fig-33mK}~(b) presents the total fluorescence
from the ODT, proportional to the number of trapped atoms, versus
the focus position. It is evident that when the centers of MOT and
ODT coincide (ODT focus position $z=0$), the number of captured
atoms is over 5 times smaller than in the case of
displaced ODT: $z=\pm10$~mm. The observed asymmetry between
positive and negative $z$ positions is attributed to geometrical
imperfections. We believe that for perfect alignment of the laser
beams and magnetic field, the $I(z)$ dependences in
Fig.~\ref{fig-33mK}~(b) should be symmetric. The difference in
heights of the two curves shown in Fig.~\ref{fig-33mK}~(b) is most
likely caused by some drift of the MOT conditions.

We have conducted the described measurement for a range of ODT
trap depths, $U_0/k\rmi{B}$, between 2.1 and 111 mK. 
For the trap depth of 2.1~mK, we obtain a
single, broad loading maximum at $z=0$. The width of this maximum
(FWHM) is 9~mm. For increasing ODT potential depth, the
maximum broadens and eventually, at 8.4 mK, splits into
two separate maxima such as shown in
Fig.~\ref{fig-33mK}~(b).

\begin{figure}
\includegraphics{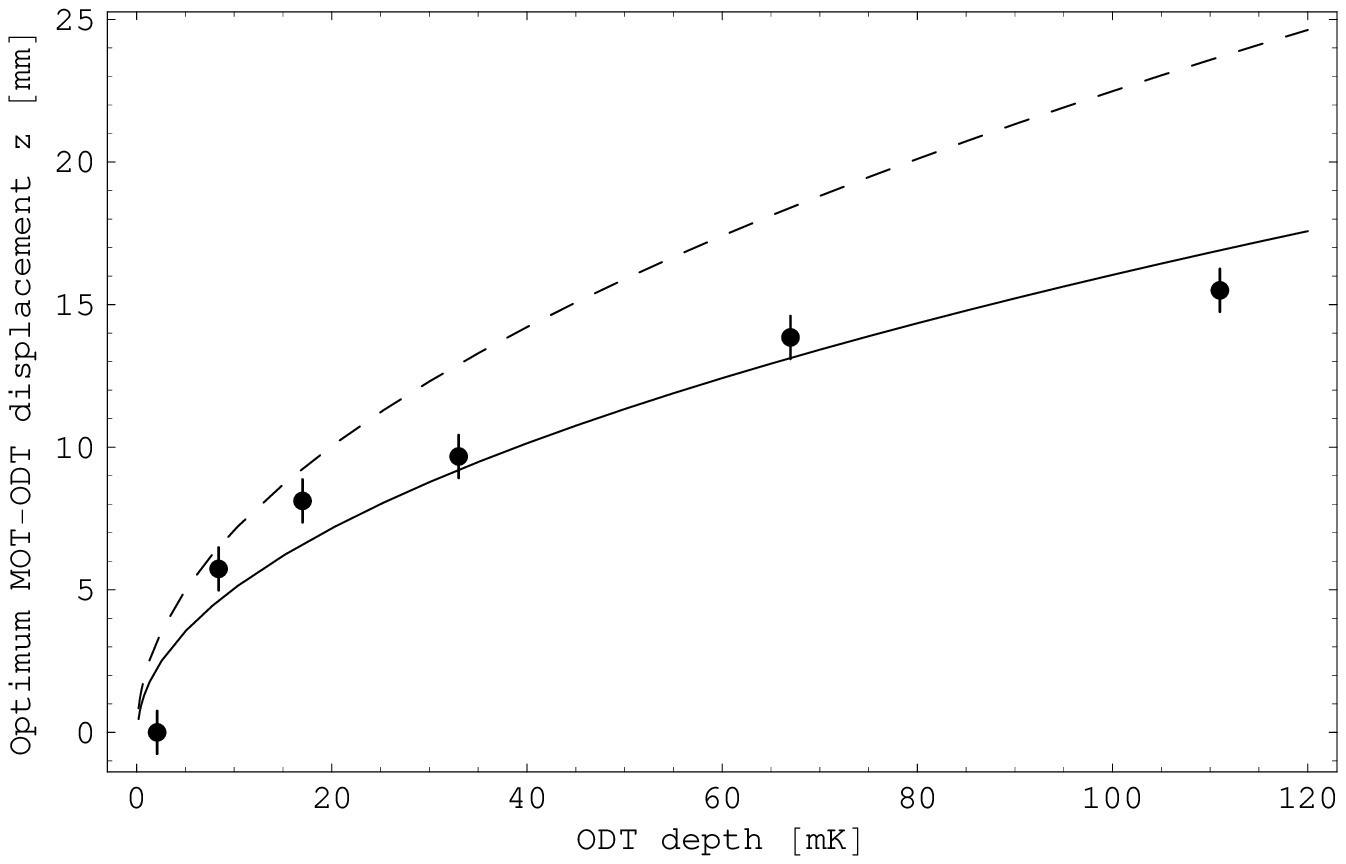}%
\caption%
{%
 \label{fig-results}%
 Optimum displacement of the optical dipole trap
 relative to the magneto-optical
 trap for maximum transfer of atoms versus the ODT depth.
 The black dots represent our experimental data.
 The solid line is calculated from Eq.~(\ref{eq-sqr-2})
 with the scaling parameter $\alpha=2.5$, as described in the text.
 The dashed line is the prediction based on the model
 of O'Hara \emph{et al.} with no
 free parameters~\cite{OHara2001}. See also Sec.~\ref{discussion}.
}
\end{figure}

With our fixed beam waist, the optimal separation of MOT and ODT
depends only on the ratio $P\rmi{ODT}/\Delta\lambda\rmi{ODT}$,
that is, it depends only on the ODT depth. We have verified this
for trap depths of 8.4, 17, and 33~mK realized by different ODT
laser powers and detunings. We have verified that optimum
displacements $z$ corresponding to a given $P\rmi{ODT} /
\Delta\lambda\rmi{ODT}$ ratio were constant within the
experimental error of a few percent for various combinations of
$P\rmi{ODT}$ and $\Delta\lambda\rmi{ODT}$ which preserved the
$P\rmi{ODT} / \Delta\lambda\rmi{ODT}$ ratio.
Figure~\ref{fig-results} presents the dependence of the
optimum displacement on the trap depths. It is worth noting that
in the case of the deepest optical traps realized with our maximum
ODT beam power, this optimum separation reaches 15 mm. This
displacement is over 10 times larger than the diameter of
the MOT, whereas the optimum displacements reported in
Ref.~\cite{Kuppens2000} were only of the order of one-half of a
MOT diameter.

\section{Discussion\label{discussion}}

The dipole potential depends on the local light intensity $I(\mathbf r)$ and detuning
from the atomic transition $\delta\lambda$ as
\begin{equation}
    \label{eq-potential}
    U\rmi{ODT}(\mathbf r)= \frac{2 \pi^2 c^3}{\omega_0^5}
    \frac{\Gamma}{\delta\lambda}I(\mathbf r),
\end{equation}
where $\Gamma$ denotes transition linewidth \cite{Kuppens2000}. The intensity of a focused Gaussian
beam of power $P$ is described by
\begin{equation}
    \label{eq-intensity}
    I(r,z)=\frac{2P}{\pi w^2(z)}\exp\left(-2\frac{r^2}{w^2(z)}\right),
\end{equation}
where $(r,z)$ are the cylindrical coordinates, and:
\begin{equation}
    \label{eq-radius}
    w(z) = w_0 \sqrt{1+(z/z\rmi{R})^2}
\end{equation}
is the $1/e^2$ beam radius. The characteristic dimension in the radial
direction is the beam waist $w_0$, while in the axial direction it is the Rayleigh length
$z\rmi{R}=\pi w_0^2/\lambda$.

\begin{figure}
\includegraphics{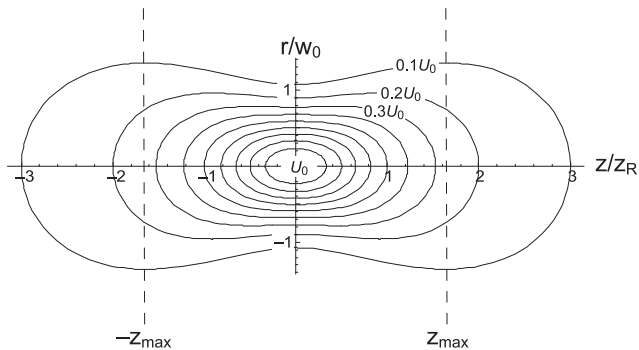}%
\caption%
{%
 \label{fig-Gaussian-beam}%
 The cross sections of the equipotential surfaces of the ODT formed by a focused Gaussian beam.
 $z/z\rmi{R}$ is the distance along the beam in units of the Rayleigh length.
 $r/w_0$ is the radial coordinate in units of the beam waist. $U_0$ is the trap depth.
 Dashed lines mark the positions of $\pm z\rmi{max}$ for $U = 0.1\,U_0$.
 }
\end{figure}

The resulting trapping potential is visualized in
Fig.~\ref{fig-Gaussian-beam}. The potential is harmonic near the
focus, with ellipsoidal equipotential surfaces. However, the more
distant equipotential surfaces are qualitatively different
from the inner ones, and acquire characteristic peanut
shape. From equations (\ref{eq-potential}) and
(\ref{eq-intensity}) one can deduce that the equipotential
surfaces change their character when $U$ becomes smaller than
$U'=U_0/e$, where $U_0$ is the trap depth and $e$ is the base of
the natural logarithm. The equipotential surfaces
$U=\mathrm{const}$ with $|U|<|U'|$ have two locations
$z\rmi{max}$ where their circumference is maximized:
\begin{equation}
    \label{eq-sqr-1}
    z\rmi{max}=\pm z\rmi{R} \sqrt{\frac{1}{e}\frac{U_0}{U}-1}.
\end{equation}

We will now compare our experimental data with the two models of
the ODT loading described in the literature
\cite{Kuppens2000,OHara2001}.

\begin{figure}
\includegraphics{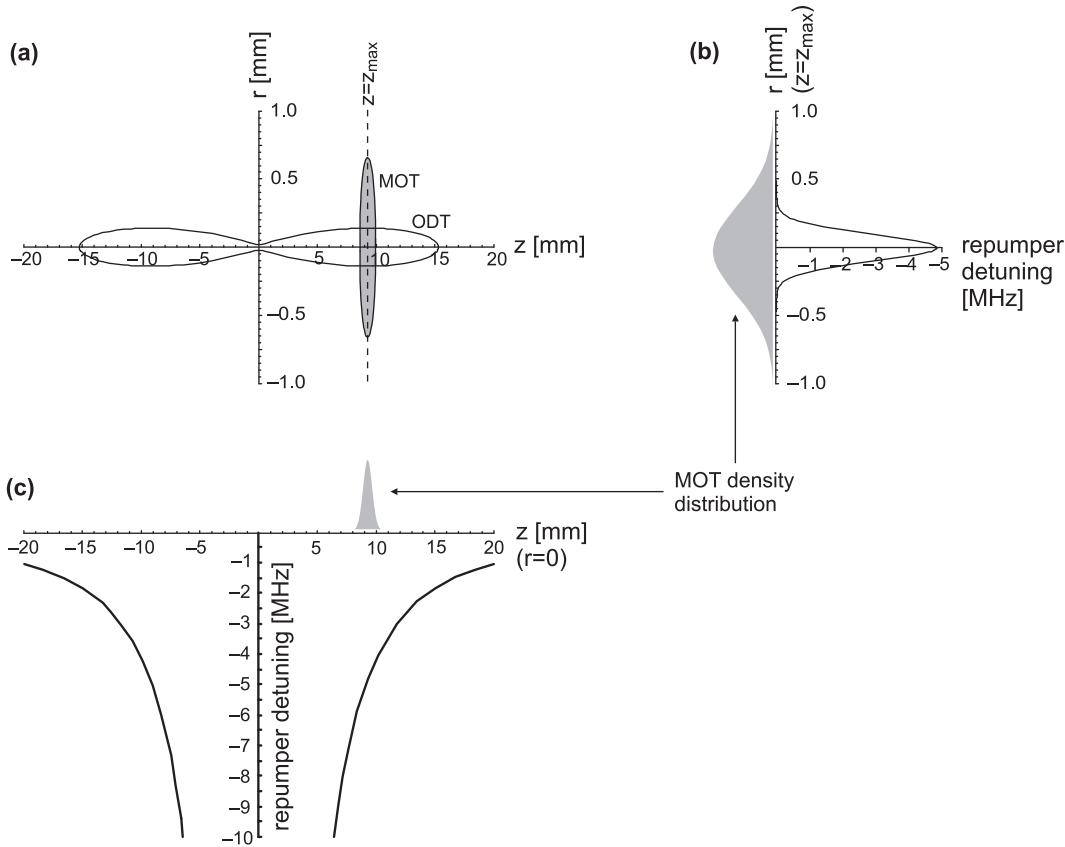}%
\caption%
{%
 \label{fig-geometry-and-detuning}
The geometrical superposition of the trap potentials and the
influence of the ODT beam on the MOT. (a)~The geometry of the MOT
(represented by the $1/e^2$ density outline) and the
33-mK-deep ODT (represented by the equipotential surface
$U=-2.5k\rmi{B}T\rmi{MOT}$ -- see text). The displacement of the
traps is optimized for maximum atom transfer. (b)~The detuning of
the MOT-repumper caused by the presence of the ODT beam,
plotted along the dashed line in (a). The gray outline represents
the atomic density distribution of the MOT. (c)~Same as (b), but
along the ODT beam axis ($r=0$). Note the vertical scale
has been expanded for better visualization.
}
\end{figure}

Kuppens \emph{et al.} \cite{Kuppens2000} assume that the initial
loading rate is proportional to the MOT atomic density
$n\rmi{MOT}$, the mean velocity of atoms $\bar v$, the effective
surface area of ODT, $A$, and the trapping probability,
$P\rmi{trap}$. The first three parameters contribute to the flux
of atoms into the ODT volume. When studying geometrical effects,
it is important to consider the effective surface area $A$. As the
Rayleigh length is much larger that the size of the MOT, the radii
of the ODT equipotential surfaces change insignificantly
over the MOT region (see also
Fig.~\ref{fig-geometry-and-detuning}(a) below). The effective
surface area across which atoms are loaded into the ODT is then
proportional to the local radius of the ODT. The crucial question
in this reasoning is: which of the equipotential surfaces should
represent the ``surface of the ODT''. Kuppens \textit{et al.} assume the
surface $U\approx k\rmi{B}T\rmi{MOT}$, where the ODT potential $U$
is comparable to the kinetic energy of atoms in the MOT.

In order to compare quantitatively our data with the model described
above, we write the equation describing the ODT surface as
$U= \alpha k\rmi{B} T\rmi{MOT}$ and look for a scaling factor $\alpha$,
of the order of unity, which gives the best fit to our data.
Equation~(\ref{eq-sqr-1}) now takes the form
\begin{equation}
\label{eq-sqr-2}
z=\pm z\rmi{R} \sqrt{\frac{1}{e}\frac{U_0}{\alpha k\rmi{B} T\rmi{MOT}}-1}.
\end{equation}
Using the least-squares method, we find that Eq.~(\ref{eq-sqr-2})
best describes our experiment for $\alpha=2.5$. The result of this
fitting is shown by the solid line in Fig.~\ref{fig-results}. It
turns out that despite its simplicity, the model by Kuppens
\emph{et al.} gives a good qualitative description of our
data, but needs the scaling factor $\alpha$ to be determined from
the experiment.

Another related work is that of O'Hara \emph{et
al.}~\cite{OHara2001} who developed a microscopic model of ODT
loading based on the Fokker-Planck equation. Their model is
capable of describing the spatial and temporal changes of the atom
density during ODT loading. For comparison with our results, the
relevant equation is the expression for the initial number of
atoms $N(0)$ contained in the trap (\cite{OHara2001}, Eqs. 11, 12)
\begin{equation}
    \label{eq-OHara11and12}
    N(0) = (\pi^{3/2} R n_0)\, \frac{w^2}{2} q^2 \int_0^1 dv\,v(-\ln v)\exp(-qv),
\end{equation}
where $w$ is the $1/e^2$ beam radius (the width of the ODT
potential in the radial direction), $R$ and $n_0$ are the radius
and the density of the MOT cloud, respectively, and $q=U_0 /
k\rmi{B} T\rmi{MOT}$. The calculations of O'Hara \emph{et al.}
were performed for the case when MOT and ODT centers coincide, but
remain valid also if the centers are displaced axially. Then $U_0$
in Eq.~(\ref{eq-OHara11and12}) is treated not as a global
trap depth $\displaystyle U_0=\min_{r,z}\{U(r,z)\}$, but as a
local trap depth for a particular axial position $z$:
\begin{equation}
    U_0(z)=\min_{r}\{U(r,z)\}=\frac{U_0}{1+(z/z\rmi{R})^2}
\end{equation}
(compare Eqs.~(\ref{eq-potential}) and (\ref{eq-intensity})). On
the other hand, the local beam radius $w(z)$ increases from its
initial value $w_0$ as one moves away from the focus, according to
Eq.~(\ref{eq-radius}). Consequently, as the axial
coordinate $z$ increases, the trap becomes shallower and
more energetic atoms remain untrapped. On the other hand,
the trap also becomes broader and atoms are captured from a
larger volume. The balance of these two competing mechanisms
results in a maximum loading for a certain displacement
between the MOT and ODT centers.

\begin{figure}
    \includegraphics{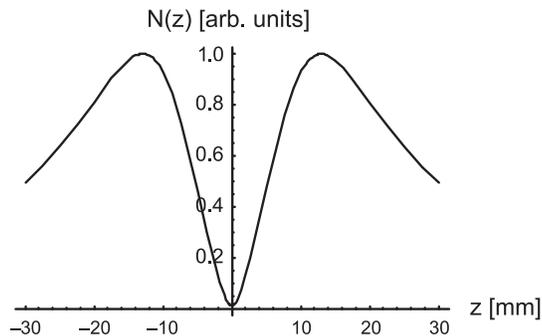}%
    \caption%
    {%
     \label{fig-OHara-z}%
     The initial number of trapped atoms, $N$,
     as a function of the MOT--ODT displacement, $z$, predicted
     by the model of O'Hara \textit{et al.}~\cite{OHara2001}
     for our experiment (trap depth = 33~mK).
    }
\end{figure}

The initial number of trapped atoms has been calculated according
to Eq. (\ref{eq-OHara11and12}) for different displacements z,
treating the parameters $w$ and $q$ as $z$-dependent. The
resulting function is plotted in Fig.~\ref{fig-OHara-z} for the
trap depth $U_0/k\rmi{B}=33$~mK. Note the maximum at $z=\pm13$~mm,
close to our experimental result: $z=\pm10$~mm (see
Fig.\ref{fig-results}). Similarly, we calculated the maxima of the
$N(z)$ function for different trap depths and the result is shown
by the dashed line in Fig.~\ref{fig-results}. It agrees
qualitatively with the experimental data, but for deeper traps the
predicted optimum displacement is overestimated. Nevertheless, the
agreement is remarkable for a model that does not require any free
parameters. It should be stressed that the model described by
O'Hara \emph{et al.} was developed for lithium atoms in
$\mathrm{CO}_2$ laser traps, where the trap-induced light
shift of the atomic transition frequency is small
\cite{OHara2001}. This is not the case of our experiment, hence
this may be the main reason of the discrepancy between
our data and the model of O'Hara \emph{et al.}

Under conditions of our experiment, the frequency shifts for the
MOT trapping and repumping transitions may be taken with good
accuracy as $2U(r,z)/h$, where $U(r,z)$ is the ODT potential which
takes into account solely the D2 line. As it was already mentioned
above, under our experimental conditions, the $^{5}P_{3/2}$
excited state makes the most important contribution to the ODT
potential and light shifts. Other states, $^{5}P_{1/2}$ and
$^{5}D_{3/2, \, 5/2}$, contribute nearly two orders of magnitude
less and can be neglected at the current level of experimental
precision. The light shift also modifies the trapping
transition but, given the large detuning (100 MHz) of the MOT
beams in the loading phase, its effect on the
trapping-beam detuning is negligible. On the other hand,
since the repumper beam is resonant, the repumping
efficiency at $z=z\rmi{max},r=0$ is reduced by about a
factor of four (for all trap depths) which allows to accumulate
atoms in the dark $F=1$ ground state. In
Fig.~\ref{fig-geometry-and-detuning}, we compare the spatial
ranges in which the light shift generates the repumper
detuning with the spatial MOT extension calculated for the trap
depth of 33~mK. As can be seen in
Figure~\ref{fig-geometry-and-detuning}(b), this reduction
affects only the central part of the MOT which,
consequently, results in a ``dark spot'' effect which is known to
increase the MOT density \cite{Ketterle1993} and enhance the trap
loading process \cite{Wieman2000}.
Figure~\ref{fig-geometry-and-detuning}(c) shows that the
light-shift-induced detuning of the repumper
beam, and consequently the dark spot effect, decrease
for increasing $z$. For $z=13$, which would be the optimum
displacement predicted from the model of O'Hara \textit{et al.},
the repumper detuning amounts only 2.5 MHz and
the dark spot effect is much weaker. Thus, the
light shift and the related dark-spot effect decrease the optimum
MOT--ODT displacement relative to the predictions of
Ref.~\cite{OHara2001}, as illustrated in Figure~\ref{fig-results}. 
We believe this is the main source of
discrepancy between our results and the model of O'Hara \textit{et
al.} which refers to the case of negligible light-shifts.

Another factor of potential importance for the ODT dynamics is the photon
scattering rate. With a trap relatively close to resonance, such as ours, the
spontaneous photon scattering leads to atom heating and, eventually, to 
atom loss. We have estimated the heating rates under conditions
of our experiment. For the deepest trap of 111~mK the scattering rate is
430~kHz and the heating during the 50~ms holding time is estimated to be about 
2~mK. This estimation was done for the trap center, whereas, as we
demonstrated, the atoms are most efficiently trapped far off the center (15mm),
where the light intensity is significantly lower
and the local scattering rate is only 530~Hz. 
Thus, we believe that it is safe to neglect this mechanism 
in the case of our experiment.

\section{Summary}

We have performed
a systematic study of the efficiency of the ODT loading from a MOT
as a function of geometrical arrangement of the ODT and MOT in a
wide range of ODT depths (2 to 110 mK). We have observed
that for the optimum loading the trap centers need to be displaced
in the direction of the ODT beam. While similar
conclusion was formulated earlier in
Ref.~\cite{Kuppens2000}, we were able to verify it in a more than
20-times wider range. We confirmed that the optimum
displacement depends only on the trap depth, i.e., on the ratio
between the trap light power and its detuning. Analyzing the
optimum displacements, we were able to quantitatively compare our
results with the existing models of Kuppens \emph{et al.}\
\cite{Kuppens2000} and O'Hara \emph{et al.}\ \cite{OHara2001}. Our
data are well described by the former model after fitting of a
scaling factor and agree qualitatively with the latter model
without any free parameters.

Based on the present work, we propose the following
semi empirical formula for the optimal trap displacement:
\begin{equation}
\label{eq-sqr-3}
z=\pm z\rmi{R} \sqrt{\frac{1}{e}\frac{U_0}{2.5 k\rmi{B} T\rmi{MOT}}-1}.
\end{equation}

\begin{acknowledgments}

This work was partly supported by Polish Ministry of Science, grant
no. NN202175835.
A.S. acknowledges also the funding from the University of Wroc\-ław, grant no. 2016/W/IFD/2005.

\end{acknowledgments}


\bibliography{displacement-3.bbl}

\end{document}